\documentstyle[epsfig,11pt]{article}

\def\beq{\begin{equation}}
\def\eeq{\end{equation}}
\def\phid{\phi^{\dagger}}
\def\noi{\noindent}

\def\talpha{\tilde{\alpha}}
\def\tbeta{\tilde{\beta}}
\def\tgamma{\tilde{\gamma}}

\begin{document}
\begin{center}
{\Large\bf Pomeron loops in the perturbative
QCD with large $N_c$ }
\vspace{0.5 cm}

M.A.Braun

{\small\it  Department of High
Energy Physics,
University of St. Petersburg,\\ 198904 St. Petersburg, Russia}\\
\end{center}
\vspace{1 cm}

\noi {\bf Abstract}

The lowest order pomeron loop is calculated for the leading
conformal weight with full dependence of the triple pomeron vertex
on intermediate conformal weights. The loop is found to be convergent.
Its contribution to the pomeron Green function begins to dominate
already at rapidities 10$\div$15. The pomeron pole renormalization is
found to be quite small due to a rapid fall of the triple pomeron vertex
with rising conformal weights.

\section{Introduction}
In the framework of QCD, in the limit of large number of colours,
strong hadronic interactions are mediated by the exchange of
BFKL pomerons which split and fuse by triple pomeron  vertices.
This picture can be conveniently described by an effective nonlocal
quantum field theory ~\cite{braun1}. A remarkable property of this theory
is its inherent conformal invariance, which is  broken by interactions with
colliding hadrons. In terms of Feynman diagrams contributions standardly
separate into tree diagrams and diagrams with pomeron loops.
For reactions with heavy nuclei the tree diagram contribution is
enhanced by factor $A^{1/3}$ for each interaction and so dominates.
This dominating part can be summed by the Balitski-Kovchegov equation
for DIS on a heavy nucleus ~\cite{bal,kov} or by a pair of equations
constructed by the author for nucleus-nucleus scattering ~\cite{braun2}.
However reactions with single hadrons require taking into account also
loop diagrams. There has been several attempts to calculate the
contribution  of a single pomeron loop ~\cite{pechan,baryva} with
a crude approximation for the triple pomeron vertex and contradicting results.
In ~\cite{baryva} it was found that the magnitude of the loop is so small
that it gives no significant contribution up to extaordinary high energies
(rapidities of the order of 40).
Lately there were several claims to sum all loop contributions in the
colour dipole approach or in the so-called reaction-diffusion formulation
of the scattering
mechanism ~\cite{lo1,lo2,lo3,lo4,lo5}. However very crude approximations
made from the start do not allow to consider these results even minimally
reliable.
These circumstances give us a motivation to reconsider the contribution of a
single pomeron loop by the conformal invariant technique developed in
~\cite{braun1}. This technique in fact greatly simplifies the
derivation and allows to fix  numerical coefficients, uncertainties in
which in our opinion were one of  the reason why the results
of ~\cite{pechan} and ~\cite{baryva} turned out to be different.
Most important we also use
the exact form of the triple pomeron vertex, which appears very different
from its approximate value used in the previous calculations.

Our results first demonstrate that the pomeron loop,
with all contributions taken into account,  is finite and does not need
renormalization, in contrast to the old local Regge-Gribov model.
Its numerical value is found to be small indeed, but not so small as
calculated in  ~\cite{baryva}. As a result, its influence becomes visible
at much lower energies than claimed there. With realistic values for
the QCD (fixed) coupling constant its contribution starts dominating already
at rapidities $y\sim 10-15$. This of course means that taking loops into
account for reactions with single hadrons is necessary already at present
energies.

The paper is organized as follows. In the next section we recall
some elements of the conformal technique introduced in
~\cite{braun1} to be used in loop calculations. Using this technique
we calculate the loop contribution in the next section. Section 4.
presents our numerical results and discusses influence of the
loop contribution on the pomeron Green function. Some conclusions are
drawn in the last section. Technical details and comparison with
~\cite{baryva} are discussed in three appendices.

\section{The  pomeron interaction diagrams}
Feynman diagrams for the pomeron interaction are built from the
pomeron propagator and triple pomeron vertex. The pomeron propagator
\beq
g_{y_1-y_2}(r^{(1)}_1,r^{(1)}_2|r^{(2)}_1,r^{(2)}_2)\equiv
g_{y_1-y_2}(1|2),
\eeq
where $1=\{r^{(1)}_1,r^{(1)}_2\}$ are the initial coordinates of the
two reggeized gluons, $2=\{r^{(2)}_1,r^{(2)}_2\}$ are
their final coordinates and $y_1-y_2$ is the rapidity difference,
satisfies the equation
\beq
\Big(\frac{\partial y}{\partial}+H\Big)
g_{y-y'}(1|2))=
\delta(y-y')\nabla_1^{-2}\nabla_2^{-2}\delta(1|2),
\eeq
where $H$ is the BFKL Hamiltonian ~\cite{lip}.
The triple pomeron vertex can be read off the interaction
Lagrangian
\beq
L_I=\frac{2\alpha_s^2N_c}{\pi}
\int\frac{d^2r_1d^2r_2d^2r_3}{r_{12}^2r_{23}^2r_{31}^2}
\phi(1)\phi(2) L_{12}\phid (3) +h.c,
\label{lagran}
\eeq
where $\phi$ and $\phid$ are the two fields which describe the
propagating pomerons,  operator$L_{12}$ is defined as
\beq
L_{12}=r_{12}^4\nabla_1^2\nabla_2^2
\eeq
and the fields in (\ref{lagran}) are to be taken at the same rapidity.

In absence of interaction with external hadrons the theory is conformal
invariant. This makes it convenient to pass to the conformal basis
formed by functions
(in complex notation) ~\cite{lip}
\beq
E_\mu(1)=E_{\mu}(r_1,r_2)
=\left(\frac{r_{12}}{r_{10}r_{20}}\right)^{h}
\left(\frac{r^*_{12}}{r^*_{10}r^*_{20}}\right)^{\bar{h}},
\label{estate}
\eeq
Here $\mu=\{n,\nu,r_0\}=\{h,r_0\}$, $h=(1-n)/2+i\nu$,
$\bar{h}=(1+n)/2+i\nu$, with $n$ integer, $\nu$ real and
two-dimensional transverse $r_0$, enumerate the basis. In the following,
for clarity, we shall  sometimes write $h$ as a set of two numbers
$\{n,\nu\}$.
We also pass from rapidity $y$ to complex angular momentum $j=1+\omega$:
\beq
g_{y}=
\int_{a-i\infty}^{a+i\infty}\frac{d\omega}{2\pi i}e^{\omega y}g_{\omega}.
\eeq
Then the propagator can be presented as
\beq
g_\omega(1|2)=\sum_{\mu>0}E_{\mu}(1)E^*_\mu(2)g_{\omega,h},
\label{sumforg}
\eeq
where
\beq
\sum_{\mu}=\sum_{n=-\infty}^{\infty}\int_0^\infty d\nu\frac{1}{a_h}
 \int d^2r_0,
\label{confsum}
\eeq
with
\beq
a_{h}\equiv a_{\mu}=\frac{\pi^4}{2}\frac{1}{\nu^2+n^2/4}.
\eeq
The conformal propagator is
\beq
g_{\omega,h}=\frac{1}{l_{n\nu}}\,\frac{1}{\omega-\omega_{h}},
\label{confprop}
\eeq
where $\omega_{h}$ are the BFKL levels
\beq
\omega_{h}=2\bar{\alpha}\Big(\psi(1)-{\rm Re}\,\psi(h)\Big),
\ \ \bar{\alpha}=\frac{\alpha_sN_c}{\pi}
\eeq
and
\beq
l_{h}=\frac{4\pi^8}
{a_{n+1,\nu}a_{n-1,\nu}}.
\eeq

The triple pomeron vertex can be presented in the
conformal basis as
\beq
\Gamma(1|2,3)=\sum_{\mu_1,\mu_2\mu_3>0}E_{\mu_1}(1)E^*_{\mu_2}(2)
E^*_{\mu_3}(3)\Gamma_{\mu_1|\mu_2\mu_3},
\label{sumforv}
\eeq
The dependence on the intermediate c.m. coordinates  $R_1$, $R_2$
and $R_3$ is fixed by conformal invariance
\beq
\Gamma_{\mu_1|\mu_2\mu_3}=R_{12}^{\alpha_{12}}R_{23}^{\alpha_{23}}
R_{31}^{\alpha_{31}}\cdot
(a.\ f.)\cdot
\Gamma_{h_1|h_2,h_3}.
\label{confver}
\eeq
Here $(a.f.)$ means  the antiholomorhic factor.
Powers $\alpha_{ik}$ are known
\[
\alpha_{12}=-\frac{1}{2}+\frac{1}{2}(n_2-n_1-n_3)+i(\nu_1-\nu_2+\nu_3),\]\[
\alpha_{23}=-\frac{1}{2}+\frac{1}{2}(n_1+n_2+n_3)-i(\nu_1+\nu_2+\nu_3),\]
\beq
\alpha_{31}=-\frac{1}{2}+\frac{1}{2}(n_3-n_1-n_2)+i(\nu_1+\nu_2-\nu_3)
\label{alphas}
\eeq
The powers in the
antiholomorhic factor $\tilde{\alpha}$'s are obtained by changing
$n_i\to -n_i$.  In the lowest approximation
the conformal vertex $\Gamma^{(0)}_{h_1|h_2,h_3}=
\Omega_{\bar{h}_1,h_2,h_3}$
with $\bar{h}=\tilde{h}^*$ was
introduced and studied by Korchemsky \cite{korch}.

Using (\ref{sumforg}), (\ref{sumforv}) and the orthonormlization
property  of states (\ref{estate}) one can perform integrations
over gluon coordinates and substitute them by summations over conformal
weights  and integration over center-of-mass coordinates, as
indicated in (\ref{confsum}). With the known expressions for the
propagator and vertex given by (\ref{confprop}) and (\ref{confver})
respectively, one can then write expressions for any Feynman diagram
directly in the conformal basis. One should only take into account that
'energies' $\omega$ are conserved at each vertex.

\section{The pomeron self-mass}
\subsection{The pomeron full Green function}
The full pomeron Green function $G_\omega(1|2)$ can also be written
in the form similar to (\ref{sumforg}):
\beq
G_\omega(1|2)=\sum_{\mu>0}E_{\mu}(1)E^*_\mu(2)G_{\omega,h}.
\label{sumforg1}
\eeq
The Schwinger-Dyson equation  expresses the full Green function
via the pomeron self-mass:
\beq
G_{\omega,h}=\frac{1}{1/g_{\omega,h}+
l_{h}^2\Sigma_{\omega,h}}.
\eeq
where $\Sigma_{\omega,h}$ is the pomeron self-mass in the
conformal basis.

Similar to (\ref{sumforg}) the pomeron self-mass in the gluon coordinate space
can be written as a sum over conformal eigenstates
\beq
\Sigma (1|1')=
\sum_{\mu_{1},\mu'_{1}}\Sigma_{\mu_{1}\mu'_{1}}
E_{\mu_{1}}(1)E^*_{\mu'_{1}}(1'),
\eeq
where the self-mass in the conformal representation is
\beq
\Sigma_{\mu_1\mu'_1}=
\frac{8\alpha_s^4N_c^2}{\pi^2}
\int\frac{d\omega_1}{2\pi i}
\sum_{\mu_2,\mu_3}\Gamma^{(0)}_{\mu_1|\mu_2,\mu_3}
G_{\mu_2}G_{\mu_3}
\Gamma_{\mu_2\mu_3|\mu'_1}.
\label{sig1}
\eeq
and the suppressed  dependence on $\omega$ is obvious from its conservation
at the vertexes.
 From its conformal invariance it follows that
\beq
\Sigma_{\mu_{1}\mu'_{1}}=\delta_{\mu_{1}\mu'_{1}}
\Sigma_{\mu_{1}},
\label{conform}
\eeq
where
\beq
\delta_{\mu\mu'}=\delta_{nn'}\delta(\nu-\nu')\delta^2(r_{00'})a_{h}.
\label{delta}
\eeq
Summation over $\mu_2$ and $\mu_3$ in (\ref{sig1}) includes integration
over two center-of-mass coordinates $R_2$ and $R_3$. Dependence on
them comes
only from the vertex functions and indicated in (\ref{confver}) and
(\ref{alphas}).

The second vertex part $\Gamma_{\mu_2\mu_3|\mu'_1}$ is defined by the
expansion
\beq
\Gamma(2,3|1)=\sum_{\mu_1,\mu_2,\mu_3>0}E^*{\mu_1}(1)E_{\mu_2}(2)E_{\mu_3}(3)
\Gamma_{\mu_2\mu_3|\mu'_1}.
\eeq
We use the property of functions $E_{\mu}$
\beq
\int d(1)E^*_{\mu'}(1)E_{\mu}(1)=\delta_{\mu'\mu}, \ \ \mu,\mu'>0,
\eeq
where $d(1)=d^2r_1d^2r'_1/r_{11'}^4$,
to find
\beq
\Gamma_{\mu_2\mu_3|\mu'_1}=\int d(1)d(2)d(3)E_{\mu_1}(1)E^*_{\mu_2}(2)E^*_{\mu_3}(3)
\Gamma(2,3|1).
\eeq
However $G(2,3|1)=G(1|2,3)$ and
\beq
E^*_{\mu}(1)=E_{\bar{\mu}}(1), \ \ \bar{\mu}=\mu(n\to -n,\nu\to -\nu),
\eeq
so that
\beq
\Gamma_{\mu_2\mu_3|\mu'_1}=\int d(1)d(2)d(3)E^*_{\bar{\mu}_1}(1)E_{\bar{\mu}_2}(2)E_{\bar{\mu}_3}(3)
\Gamma(1|2,3)=\Gamma_{\bar{\mu}_1|\bar{\mu}_2\bar{\mu}_3}.
\eeq

So we find an integral in (\ref{sig1})
\[
I(R_1,R'_1,h_1,h'_1)\]\beq=
\int d^2R_2d^2R_3
R_{12}^{\alpha_{12}}R_{23}^{\alpha_{23}}R_{31}^{\alpha_{31}}
{R_{12}^*}^{\tilde{\alpha}_{12}}{R_{23}^*}^{\tilde{\alpha}_{23}}{R_{31}^*}^{\tilde{\alpha}_{31}}
R_{1'2}^{{\beta}_{1'2}}R_{23}^{{\beta}_{23}}R_{31'}^{{\beta}_{31'}}
{R_{1'2}^*}^{\tilde{\beta}_{1'2}}{R_{23}^*}^{\tilde{\beta}_{23}}
{R_{31'}^*}^{\tilde{\beta}_{31'}}.
\label{integ1}
\eeq
The full set of powers is
\[
\alpha_{12}=-\frac{1}{2}+\frac{1}{2}(n_2-n_1-n_3)+i(\nu_1-\nu_2+\nu_3),\]\[
\alpha_{23}=-\frac{1}{2}+\frac{1}{2}(n_1+n_2+n_3)+i(-\nu_1-\nu_2-\nu_3),\]
\[
{\alpha}_{31}=-\frac{1}{2}+\frac{1}{2}(n_3-n_1-n_2)+i(\nu_1+\nu_2-\nu_3)
\]
\[
\tilde{\alpha}_{12}=-\frac{1}{2}+\frac{1}{2}(-n_2+n_1+n_3)+i(\nu_1-\nu_2+\nu_3),\]\[
\tilde{\alpha}_{23}=-\frac{1}{2}+\frac{1}{2}(-n_1-n_2-n_3)-i(\nu_1+\nu_2+\nu_3),\]
\[
\tilde{\alpha}_{31}=-\frac{1}{2}+\frac{1}{2}(-n_3+n_1+n_2)+i(\nu_1+\nu_2-\nu_3)
\]
\[
\beta_{1'2}=-\frac{1}{2}+\frac{1}{2}(-n_2+n'_1+n_3)+i(-\nu'_1+\nu_2-\nu_3),\]\[
\beta_{23}=-\frac{1}{2}+\frac{1}{2}(-n'_1-n_2-n_3)+i(\nu'_1+\nu_2+\nu_3),\]
\[
\beta_{31'}=-\frac{1}{2}+\frac{1}{2}(-n_3+n'_1+n_2)+i(-\nu'_1-\nu_2+\nu_3)
\]
\[
\tilde{\beta}_{1'2}=-\frac{1}{2}+\frac{1}{2}(n_2-n'_1-n_3)+i(-\nu'_1+\nu_2-\nu_3),\]\[
\tilde{\beta}_{23}=-\frac{1}{2}+\frac{1}{2}(n'_1+n_2+n_3)+i(\nu'_1+\nu_2+\nu_3),\]
\beq
\tilde{\beta}_{31'}=-\frac{1}{2}+\frac{1}{2}(n_3-n'_1-n_2)+i(-\nu'_1-\nu_2+\nu_3)
\eeq

So we find
\beq
\Sigma_{\mu_1\mu'_1}=
\frac{8\alpha_s^4N_c^2}{\pi^2}
\int\frac{d\omega_1}{2\pi i}
\sum_{h_2,h_3}\Gamma^{(0)}_{h_1|h_2,h_3}
G_{\omega_1,h_2}G_{\omega-\omega_1,h_3}
\Gamma_{h_2h_3|h'_1}I(R_1,R'_1,h_1,h'_1).
\label{sig2}
\eeq
According to conformal invariance property (\ref{conform}) this
expression has to be proportional to $\delta_{\mu_1\mu'_1}$ and this has
to be valid with any
values for $G_h$ and $\Gamma_{h_1|h_2h_3}$,
since with any values for these quantities conformal
invariance of the Green and vertex functions is fulfilled.
Taking unity for the vertex functions
and deltas for the Green functions in the conformal basis
we find that the integral
(\ref{integ1}) itself has to be proportional to
$\delta_{\mu_1\mu'_1}$ and independent of $R_1$
\beq
I(R_1,R'_1,h_1,h'_1)=\delta_{n_1n'_1}\delta(\nu_1-\nu'_1)
\delta^2(R_{11'})
F(h_1|h_2,h_3).
\label{integ2}
\eeq
The pomeron self-mass is expressed via $F$ according to (\ref{conform})
and (\ref{delta}):
\beq
a_{h_1}\Sigma_{h_1}=
\frac{8\alpha_s^4N_c^2}{\pi^2}
\int\frac{d\omega_1}{2\pi i}
\sum_{h_2,h_3}\Gamma^{(0)}_{h_1|h_2,h_3}
G_{\omega_1,h_2}G_{\omega-\omega_1,h_3}
\Gamma_{h_2h_3|h'_1}F(h_1|h_2h_3).
\label{sig3}
\eeq

\subsection{The lowest order pomeron self mass
in the conformal representation}

Calculation of $F(h_1|h_2h_3)$ is described in Appendix 1.
It is found that
\beq
F(h_1|h_2,h_3)=a_{h_1}.
\label{ffun}
\eeq
This gives for the self-mass
\beq
\Sigma_{\omega,h_1}=
\frac{8\alpha_s^4N_c^2}{\pi^2}\int\frac{d\omega_1}{2\pi i}
\sum_{h_2,h_3}\Gamma^{(0)}_{h_1|h_2,h_3}G_{\omega_1,h_2}G_{\omega-\omega_1,h_3}
\Gamma_{\bar{h_1}|\bar{h_2},\bar{h_3}}.
\label{smassgen}
\eeq
Here
\beq
\sum_{h}=\frac{2}{\pi^4}\sum_{n=-\infty}^{+\infty}\int_0^{\infty}
\left(\nu^2+\frac{n^2}{4}\right)d\nu.
\eeq

In the lowest order we substitute the Green functions $G$ by
pomeron propagators $g$ given by (\ref{confprop}) and the full vertex
$\Gamma$ by its lowest order expression $\Gamma^{(0)}$.
Introducing the explicit expression for $l_h$
and doing the integral over $\omega_1$ we then obtain
\beq
\Sigma_{\omega,h}=\frac{\alpha_s^4 N_c^2}{8\pi^{10}}
\sum_{n_1,n_2=-\infty}^{+\infty}
\int_0^{\infty}
d\nu_1 d\nu_2
\frac{|\Gamma^{(0)}_{h|h_1,h_2}|^2}
{\omega-\omega_{h_1}-
\omega_{h_2}}b_{h1}b_{h_2},
\label{smass}
\eeq
where
\beq
b_h=b_{n,\nu}=
\frac{\nu^2+n^2/4}{[\nu^2+(n+1)^2/4]
[\nu^2+(n-1)^2/4]}.
\label{bdef}
\eeq

One observes that the contributions from
conformal weights $n_1=\pm 1$ and $n_2=\pm 1$ seemingly diverge at small
$\nu_1$ and $\nu_2$, when in the denominators
we find factors $\nu_1^2$ or $\nu_2^2$.
However one can demonstrate that at least for symmetric state $\{n,\nu\}$
(that is with even $n$) the vertex cannot be coupled to any of antisymmetric
states $\{n_1,\nu_1\}$ and $\{n_2\nu_2\}$ (that is with odd $n_1$ or $n_2$)
(see Appendix 2.), so that this divergence is in fact absent.
The situation for antisymmetric initial state $\{n,\nu\}$ is not clear,
since the 3-pomeron vertex was derived only for a symmetric state
($q\bar{q}$-loop). In the following we assume the initial state
$\{n,\nu\}$ to be symmetric ($n$ even).

In fact the lowest order vertex function in the conformal representation
$\Gamma^{(0)}_{h|h_1,h_2}=
\Omega(\bar{h},h_1,h_2)$ was studied in ~\cite{korch}. It was found to
be highly complicated. It was expressed in ~\cite{korch} in terms of the
Meier function $G^{pq}_{44}$. In ~\cite{korch, BNP} only its value for the
leading conformal weights
$h=h_1=h_2=1/2$ was found:
\beq
\Gamma^{(0)}_{\frac{1}{2},\frac{1}{2},\frac{1}{2}}=
\Omega\Big(\frac{1}{2},\frac{1}{2},\frac{1}{2})=
7766.679.
\eeq

Self-mass (\ref{smass}) is obviously  an analytic function of $\omega$
with a left-hand cut. For a particular term in the sum over $n_1$ and $n_2$
the cut goes from $\omega_{n_1,0}+\omega_{n_2,0}$ to $-\infty$.
The rightmost cut corresponds to $n_1=n_2=0$ and starts at $\omega=2\Delta$
where $\Delta=4\bar{\alpha}\ln 2$ is the BFKL intercept.
The discontinuity of $\Sigma$ across the cut is given by
\[
{\rm Disc}\,\Sigma_{\omega,h}=
\Sigma_{\omega+i0,h}-\Sigma_{\omega-i0,h}\]\beq=
-i\frac{\alpha_s^4 N_c^2}{4\pi^9}\sum_{n_1,n_2=-\infty}^{+\infty}
\int_0^{\infty}d\nu_1 d\nu_2
|\Gamma^{(0)}_{h|h_1,h_2}|^2b_{h_1}b_{h_2}
\delta(\omega-\omega_{h_1}-
\omega_{h_2}).
\label{disc}
\eeq
It is negative imaginary and
 different from zero for $\omega<2\omega_{1/2}=8\bar{\alpha}\ln 2$.
Therefore the original
leading BFKL pole at $\omega=\omega_{h=1/2}$ acquires an
imaginary part and splits into two complex conjugate poles
which remain on the physical sheet of the complex
$\omega$-plane in contrast to normal theories where the pole goes
under the cut onto the unphysical sheet.

It is not difficult to find the asymptotic behaviour of the
Green function in the conformal representation as a function of
rapidity
\beq
G_{y,h}=\int \frac{d\omega}{2\pi i}e^{y\omega}G_{\omega,h}.
\eeq
At $y\to\infty$ it is dominated by the contribution from the
rightmost cut extending from $\omega=2\omega_{1/2}\equiv 2\Delta$
 to $\omega=-\infty$
\beq
G_{y,h}\sim -\frac{1}{2\pi i}\int_{-\infty}^0d\omega e^{y\omega}
\frac{1}{l_{h}}{\rm Disc}\,
\frac{1}{\omega-\omega_{h}+l_{h}\Sigma_{\omega,h}}=
\frac{1}{2\pi i}\int_{-\infty}^0d\omega  e^{y\omega}
\frac{{\rm Disc}\,\Sigma_{\omega,h} }
{|\omega-\omega_{h}+l_{h}\Sigma_{\omega,h}|^2}.
\eeq
The leading contribution comes from conformal weights
$n_1=n_2=0$ and small $\nu_1$ and $\nu_2$. So we get
\beq
G_{y,0,\nu}\sim
-\frac{\alpha_s^4 N_c^2}{8\pi^{10}\Delta^2}
\int_{-\infty}^0d\omega e^{y\omega}
\int_0^{\infty}d\nu_1 d\nu_2\nu_1^2\nu_2^2
\frac{|\Gamma^{(0)}_{0\nu|0\nu_1,0,\nu_2}|^2
\delta(\omega-\omega_{0,\nu_1}-\omega_{0,\nu_2})}
{(\nu_1^2+1/4\Big)^2(\nu_2^2+1/4\Big)^2},
\eeq
where we used that at small $\nu$ $l_{n\nu}\simeq 1$.
Expanding at small $\nu_1,\nu_2$ in the standard manner
\beq
\omega_{0,\nu_{1(2)}}=\Delta-a\nu_{1(2)}^2,\ \  a=14\zeta(3)\bar{\alpha}
\eeq
and doing the integral over $\omega$ we get
\beq
G_{y,0,\nu}\sim
-\frac{\alpha_s^4 N_c^2}{8\pi^{10}\Delta^2}
e^{2y\Delta}
\int_0^{\infty}d\nu_1 d\nu_2\nu_1^2\nu_2^2
e^{-ya(\nu_1^2+\nu_2^2)}
\frac{|\Gamma^{(0)}_{0\nu|0\nu_1,0,\nu_2}|^2}
{(\nu_1^2+1/4\Big)^2(\nu_2^2+1/4\Big)^2}.
\eeq
Taking all factors which have finite values at $\nu_1=\nu_2=0$ out of
the integral we find finally
\beq
G_{y,0,\nu}\sim
-\frac{2\bar{\alpha}^4}{\pi^{5}\Delta^2 N_c^2}
\Omega^2\Big(\frac{1}{2},\frac{1}{2},\frac{1}{2})
e^{2y\Delta}\frac{1}{(ay)^3}=-Ce^{2y\Delta}\frac{1}{y^3}.
\eeq
The asymptotic is negative and of the order $1/N_c^2$ as expected.
So in principle it belongs to a higher order in the expansion in $1/N_c$.

For $N_c=3$ and taking $\alpha_s=0.2$ we have
\beq
\Delta=0.530,\ \ a=3.21,\ \ C=6.26.
\label{barep}
\eeq
With these values the Green function with a loop becomes greater
than the bare one already at $y\sim 10$.
To compare, in  ~\cite{baryva} the contribution from
the pomeron self-mass was found to be very small due to partly a smaller
numerical factor (see Appendix 3.), but mainly due to a different manner
of studying the asymptotic. The authors of ~\cite{baryva}
assumed that  the propagators around
the loop are also in the asymptotical regime and accordingly restricted
integration over $\nu$ to small values. This introduced additional
damping of the contribution due to weight $\nu^2$. With their value of the
coupling constant  $\alpha_s=0.3$ they found that the loop influence
becomes significant only at rapidities greater than 40.
However in fact the propagators around the loop enter not in their
asymptotical regime, which greatly enhances the magnitude of the loop
contribution.

To see the influence of the loop on the position of the pole in the
pomeron Green function (its 'mass renormalization') one has to calculate
the self mass in the vicinity of the pole. Restricting to the Green function
for the leading conformal weight $h=1/2$ and taking into account
in the sum (\ref{smass}) only the
leading intermediate conformal weights with $n_1=n_2=0$ one has to evaluate
$\Sigma_{\omega,1/2}$ as a function of energy $\omega$
far from the tip of the cut at
$\omega=2\Delta$. Then
one has to know $\Omega(1/2,h_1,h_2)$ as a function of
$h_1=1/2+i\nu_1$
and $h_2=1/2+i\nu_2$ at $\nu_1$ and $\nu_2$ greater than zero.
In previous calculations ~\cite{baryva}
a very crude estimate
of $\Sigma_{\omega,1/2}$ was obtained assuming that
$\Omega$ does not substantially change in the essential
integration region and can be
approximated by the known $\Omega(1/2,1/2,1/2)$.
If one uses our numerical coefficient in the expression for the self-mass
then with this approximation the real and imaginary parts of
$\Sigma_{\omega,1/2}$ at $\omega=\Delta$ turn out
to be of the same order as
the BFKL intercept $\Delta$.
The two complex conjugate poles corresponding to the "physical"
pomeron  are then found to be significantly different from the
bare pomeron pole :
\beq
\omega_P^{\pm}=0.473\pm 0.027\ i
\label{physp}
\eeq
(recall that the "bare" intercept was real and equal to $\Delta=0.530$).
The real part of the intercept is diminished in accordance with the
conclusions in  ~\cite{baryva}) although this
change is not too small due to a greater coefficient in our self-mass.
The important fact is that the pole acquires an
imaginary part of the same order as the change in the real part. This
fact challenges our standard renormalization methods, since it cannot be
compensated by adding extra terms to the original Lagrangian.

However this calculation  obviously overestimates $\Sigma$ at values
of $\omega$ far
from $\omega=2\Delta$, since in fact the three pomeron vertex function
$\Omega$ rapidly diminishes in this region.
Calculation of $\Omega(1/2,h_1,h_2)$ for different $h_{1,2}=1/2+i\nu_{1,2}$
requires a complicated numerical procedure.
Wee used the formulas derived by Korchemsky in ~\cite{korch}
in the form of integrals over variable $x$ in the interval $[0,1]$.
\footnote{
We have checked that all of them  are indeed expressed by the
Meijer function as indicated in ~\cite{korch} except
$\bar{J_1}$. For the latter the expression in
terms of the Meijer function probably contains a misprint.}
The found $\Omega(1/2,h_1,h_2)$ rapidly diminishes with $\nu_1$ and $\nu_2$.
In Fig.1 we illustrate this behaviour  showing
$|\Omega(1/2,1/2+i\nu_1,1/2+i\nu_2)|$ as a function of $\nu_2$ at different
$\nu_1$.

\begin{figure}
\hspace*{1 cm}
\epsfig{file=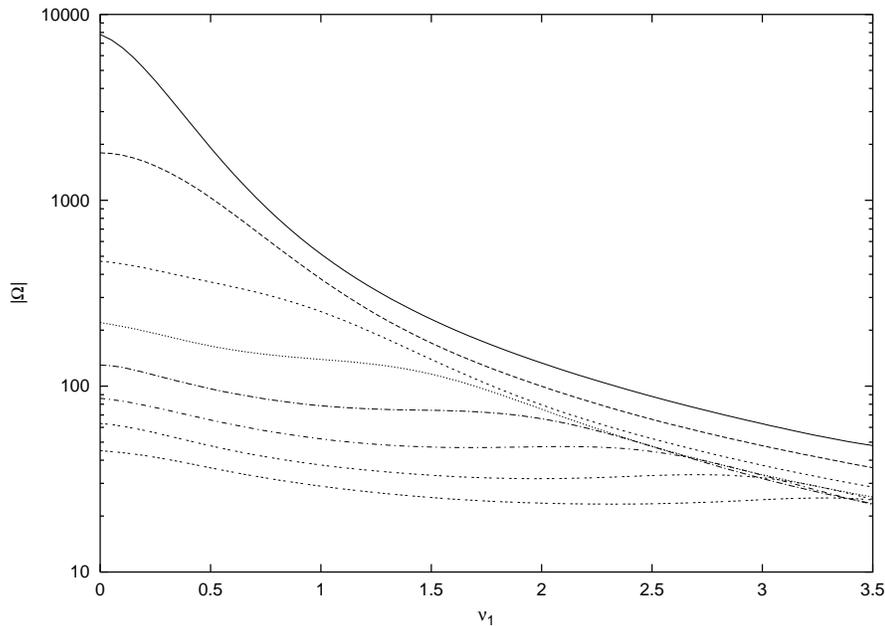,width=12 cm}
\caption{Absolute values of $\Omega(1/2,1/2+i\nu_1,1/2+i\nu_2)$
as a function of $\nu_1$ at different $\nu_2$. Curves from top to
bottom correspond to $\nu_2=0.0,0.5,1.0,1.5,2.0,2.5,3.0,3.5$.}
\label{fig1}
\end{figure}

\begin{figure}
\hspace*{4 cm}
\epsfig{file=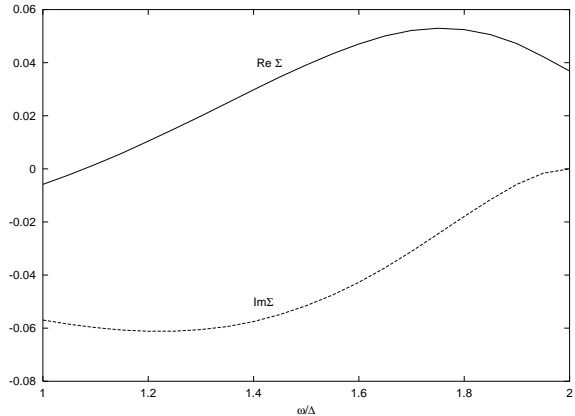,width=8 cm}
\caption{Real and imaginary parts of $\Sigma_{\omega,1/2}$
as a function of $\omega/\Delta$ in the region
$\Delta\leq\omega\leq 2\Delta$ for $\alpha_s=0.2$,
$\Delta=0.530$, with the 3-Pomeron vertex depending
on internal conformal weights}
\label{fig2}
\end{figure}

Values of
$\Sigma_{\omega,1/2}$ given by Eq. (\ref{smass}) with $\Omega(1/2,h_1,h_2)$
depending on $h_1$ and $h_1$ are shown in Fig. 3 as a function of $\omega$.
One observes that at $\omega=\Delta$ the self-mass
becomes quite small. With $\alpha_s=0.2$
\beq
\Sigma_{\Delta,1/2}=-0.0058-i0.0569,
\eeq
The renormalization of the pomeron intercept
is thus insignifiant:
the two complex conjugate poles corresponding to the "physical" pomeron are
now
\beq
\omega_P^{\pm}=0.524\pm 0.057\,i
\eeq
to be compared with the estimate (\ref{physp}), which neglects the
vertex dependence on $\nu$.

\section{Conclusions}
We have calculated the lowest order loop contribution to the pomeron
Green function in the conformal technique. The main novelty of our
calculation is the use of the triple pomeron vertex with full dependence on the
intermediate conformal weights. On the technical side,
our loop is found to carry a much greater
numerical coefficient as compared with the old calculations in
~\cite{baryva}. As  a result, the loop is not at all so inocuous
as stated in ~\cite{baryva}: its contribution begins to dominate
already at rapidities of the order 10$\div$15. On the other hand,
due to the variation of the triple pomeron vertex,
its influence on the pomeron pole (mass renormalization) is found
to be quite small in agreement with ~\cite{baryva}.

On the theoretical side we have seen that the pomeron self mass
is finite, at least in the lowest order, so that mass renormalization is
not obligatory, unlike the old local Regge-Gribov pomeron model.
We have also confirmed that due to the wrong sign of in front of
the self-mass the bare pomeron pole splits into two
complex conjugate ones, which stay
on the physical sheet, contrary to what happens in the 'normal' theory.

Finally we stress that we have limited ourselves to the lowest order
loop. With a small coupling constant this enables us to study the
asymptotic of the behaviour of the Green function only up to a certain
finite (large) rapidity determined by the condition
$\alpha_s\exp \Delta y\sim 1$, that is
\[ y<\frac{1}{\Delta}\ln\frac{1}{\alpha_s}. \]
At larger values of $y$ loops of higher order step in. The true
asymptotic at $y\to\infty$ obviously requires summation of all loops.
In our opinion, the path to achieve this goal is still
quite long.

\section{Appendix 1. Calculation of $F(h_1|h_2,h_3)$}
Using representation (\ref{integ2}) we are going to obtain $F$ by
putting $n'_1=n_1$ and integrating over  $R_1$.
Since integral (\ref{integ1}) depends only on the difference
$R_{11'}$, we can put $R'_1=0$. After that, with $n'_1=n_1$ the
integral takes the form
\[
I(R_1,n_1,\nu_1,\nu'_1)\]\beq=
\int d^2R_2d^2R_3
R_{12}^{\alpha_{12}}R_{31}^{\alpha_{31}}
{R_{12}^*}^{\tilde{\alpha}_{12}}
{R_{31}^*}^{\tilde{\alpha}_{31}}R_{23}^{\alpha_{23}+\beta_{23}}
{R_{23}^*}^{\tilde{\alpha}_{23}+\tilde{\beta}_{23}}
R_{2}^{{\beta}_{1'2}}R_{3}^{{\beta}_{31'}}
{R_{2}^*}^{\tilde{\beta}_{1'2}}{R_{3}^*}^{\tilde{\beta}_{31'}}.
\label{integ3}
\eeq

Integration over $R_1$ leads to an integral
\beq
J(R_{23})=\int d^2R_1R_{12}^{\alpha_{12}}R_{31}^{\alpha_{31}}
{R_{12}^*}^{\tilde{\alpha}_{12}}
{R_{31}^*}^{\tilde{\alpha}_{31}}=
\int d^2R_1R_{1}^{\alpha_{31}}R_{10}^{\alpha_{12}}
{R_{1}^*}^{\tilde{\alpha}_{31}}{R_{10}^*}^{\tilde{\alpha}_{12}},\ \
R_0=R_{23}.
\label{intr1}
\eeq
Its dependence on $R_{23}$ can be esily found by rescaling
$R_1=R_0z$:
\beq
J(R_{23})=R_{23}^{-n_1+2i\nu_1}{R_{23}^*}^{n_1+2i\nu_1}j(h_1,h_2,h_3),
\label{defj}
\eeq
where
\beq
j(h_1|h_2,h_3)=\int d^2zz^{\alpha_{31}}(1-z)^{\alpha_{12}}
{z^*}^{\tilde{\alpha}_{31}}(1-z^*)^{\tilde{\alpha}_{12}}.
\label{jfun}
\eeq

Putting this result into (\ref{integ3}) we find
\[
\int d^2R_1I(R_1,n_1,\nu_1,\nu'_1)\]\beq=
j(h_1|h_2,h_3)\int d^2R_2d^2R_3
R_{23}^{\gamma_{23}}
{R_{23}^*}^{\tilde{\gamma}_{23}}
R_{2}^{{\beta}_{1'2}}R_{3}^{{\beta}_{31'}}
{R_{2}^*}^{\tilde{\beta}_{1'2}}{R_{3}^*}^{\tilde{\beta}_{31'}},
\label{integ4}
\eeq
where
\beq
\gamma_{23}=-1-n_1+i(\nu'_1+\nu_1),\ \
\tilde{\gamma}_{23}=-1+n_1+i(\nu'_1+\nu_1).
\label{gamma23}
\eeq

To calculate (\ref{integ4}) we use a  Fourier transform:
\beq
R^{\alpha}{R^*}^{\talpha}=g(\alpha,\talpha)
\int \frac{d^2q}{(2\pi)^2}q^{-1-\talpha}{q^*}^{-1-\alpha}
e^{i(q^*R+qR^*)/2},
\label{rfour}
\eeq
where
\beq
g(\alpha,\talpha)=
{2\pi} i^{\alpha-\talpha} 2^{1+\alpha+\talpha}
\frac{\Gamma(1+\talpha)}{\Gamma(-\alpha)}.
\label{galpha}
\eeq
The inverse transformation is
\beq
q^{\beta}{q^*}^{\tbeta}=
\tilde{g}(\beta,\tbeta)\int d^2R
R^{-1-\tbeta}{R^*}^{-1-\beta}
e^{-i(q^*R+qR^*)/2},
\label{inverse1}
\eeq
where
\beq
\tilde{g}(\beta,\tbeta)=g^{-1}(-1-\tbeta,-1-\beta)=
\frac{1}{2\pi}(-i)^{\beta-\tbeta} 2^{1+\beta+\tbeta}
\frac{\Gamma(1+\tbeta)}{\Gamma(-\beta)}.
\label{gbeta}
\eeq
Of course
\beq
\int \frac{d^2q}{(2\pi)^2}e^{i(q^*R+qR^*)/2}=\delta^2(R).
\eeq

Using (\ref{rfour}) we present products $R_{23}^{\gamma_{23}}
{R_{23}^*}^{\tilde{\gamma}_{23}}$, $R_{2}^{{\beta}_{1'2}}
{R_{2}^*}^{\tilde{\beta}_{1'2}}$ and
$R_{3}^{{\beta}_{31'}}{R_{3}^*}^{\tilde{\beta}_{31'}}$
as Fourier transforms to obtain
\[
\int d^2R_1I(R_1,n_1,\nu_1,\nu'_1)=d(h_1,h_2,h_3)
\int \prod_{i=1}^3\frac{d^2q_i}{(2\pi)^3}d^2R_2d^2R_3
q_1^{-1-\tgamma_{23}}{q^*_1}^{-1-\gamma_{23}}\]\beq
q_2^{-1-\tbeta_{21'}}{q^*_2}^{-1-\beta_{21'}}
q_3^{-1-\tbeta_{31'}}{q^*_3}^{-1-\beta_{31'}}
e^{-i(q_1R^*_{23}+q^*_1R_{23})/2{+i(q_2R^*_{2}+q^*_2R_{2})/2}
-i(q_3R^*_{3}+q^*_3R_{3})/2},
\label{integ9}
\eeq
where
\beq
d(h_1,h_2,h_3)=j(h_1|h_2,h_3)
g(\gamma_{23}\tgamma_{23})g(\beta_{21'}\tbeta_{21'})
g(\beta_{31'}\tbeta_{31'}).
\label{dgen}
\eeq
Integrations over $R_2$ and $R_3$ give
\[(2\pi)^4\delta^2(q_1-q_2)\delta^2(q_1-q_3)\]
and we find
\beq
\int d^2R_1I(R_1,n_1,\nu_1,\nu'_1)=d(h_1,h_2,h_3)
\int\frac{d^2q}{(2\pi)^2}
q^{-3-\tgamma_{23}-\tbeta_{21'}-\tbeta_{31'}}
{q^*}^{-3-\gamma_{23}-\beta_{21'}-\beta_{31'}}.
\label{integ10}
\eeq
We have
\[
-3-\gamma_{23}-\beta_{21'}-\beta_{31'}=
-3-\tgamma_{23}-\tbeta_{21'}-\tbeta_{31'}=
-1-i(\nu_1-\nu'_1).
\]
The integral gives
$ \frac{1}{2}\delta(\nu-\nu')$,
so that we are left with the calculation of factor $d(h_1,h_2,h_3)$
defined by (\ref{dgen}) which reduces to the calculation of $j(h_1|h_2,h_3)$
defined by (\ref{jfun}).

Presenting  $R_{1}^{\alpha_{31}}
{R_{1}^*}^{\talpha_{31}}$ and $R_{10}^{\alpha_{21}}
{R_{10}^*}^{\talpha_{21}}$ in integral (\ref{intr1}) as Fourier transforms
we get
\[
J(R_0)=g(\alpha_{21}\talpha_{21})g(\alpha_{31}\talpha_{31})
\]\beq\int \frac{d^2q_1d^2q_2}{(2\pi)^4}
d^2R_1q_1^{-1-\talpha_{31}}{q^*_1}^{-1-\alpha_{31}}
q_2^{-1-\talpha_{21}}{q^*_2}^{-1-\alpha_{21}}
e^{i(q_1R^*_{1}+q^*_1R_{1})/2-i(q_2R^*_{10}+q^*_1R_{10})/2}.
\eeq
Integration over $R_1$ gives $(2\pi)^2\delta^2(q_1-q_2)$ and
we are left with
\beq
J(R_0)=g(\alpha_{21}\talpha_{21})g(\alpha_{31}\talpha_{31})I_3,
\eeq
where
\beq
I_3=
\int \frac{d^2q_1}{(2\pi)^2}
d^2R_1q_1^{-2-\talpha_{21}-\talpha_{31}}{q^*_1}^{-2-\alpha_{21}-\alpha_{31}}
e^{i(q_1R^*_{0}+q^*_1R_{0})/2}.
\eeq
According to (\ref{rfour})and (\ref{gbeta}) integration over $q_1$ gives
\beq
I_3=R_0^{-n_1+2i\nu_1}{R^*_0}^{n_1+2i\nu_1}
\tilde{g}(-1-n_1-2i\nu_1,-1+n_1-2i\nu_1),
\eeq
which means
\beq
j(h_1|h_2,h_3)=g(\alpha_{21}\talpha_{21})g(\alpha_{31}\talpha_{31})
\tilde{g}(\gamma^*_{23},\tgamma^*_{23}),
\eeq
where we used (\ref{gamma23}).

Collecting our results we find
\beq
F(h_1|h_2,h_3)=
\frac{1}{2} g(\alpha_{21},\talpha_{21})g(\alpha_{31}\talpha_{31})
\tilde{g}(\gamma^*_{23},\tgamma^*_{23})
g(\gamma_{23}\tgamma_{23})g(\beta_{21'}\tbeta_{21'})
g(\beta_{31'}\tbeta_{31'}),
\eeq
which after trivial  calculations simplifies to
\beq
F(h_1|h_2,h_3)=
\frac{\pi^4}{2(\nu_1^2+n_1^2/4)}.
\label{ffun1}
\eeq
that is to (\ref{ffun}).


\section{Appendix 2. Coupling of the 3-pomeron vertex to pomerons with
different parity}
 In our paper ~\cite{brava} we considered coupling of the 3-pomeron vertex
 to pomeron states symmetric under the interchange of gluon coordinates
(of positive parity). However in loops the intermediate pomerons in
principle may be of different parity, both positive and negative. So we have
to know how they couple to the 3-pomeron vertex.

To this end we have to recall that originally the vertex (coupled to
the loop) consists of 4 terms which differ by permutation of gluon
momenta or coordinates, so that the vertex as a whole is symmetric
in all 4 gluons which couple to outgoing pomerons.
In the momentum space we had the vertex coupled to the loop as
\beq
V(1,2,3,4)=(1/2)g^2
\Big(G(1,2+4,3)+G(1,2+3,4)+G(2,1+4,3)+G(2,1+3,4)\Big).
\label{vexpr}
\eeq
where $G$ was a known function.

We couple it to two outgoing pomerons with wave functions
$\Psi_1(1,2)\Psi_2(4,3)$.
In ~\cite{brava} passing to the coordinate space  we in fact considered only
the first term arguing that the rest will give the same for symmetric $\Psi(1,2)$
and $\Psi(3,4)$.
Now we  repeat this derivation with all 4 terms in (\ref{vexpr}) assuming that
\beq
\Psi_1(2,1)=P_1\Psi(1,2),\ \ \Psi_2(4,3)=P_2\Psi_2(3,4),\ \ P_{1,2}=\pm 1.
\eeq

We recall that the first term in (\ref{vexpr}) leads to the following
triple pomeron contribution
\beq
T^{(1)}=-\frac{1}{4}\frac{g^4N}{4\pi^3}\int_0^Y dy
\int \frac{d^2r_1d^2r_2d^2r_3}{r_{13}^2r_{12}^2r_{23}^2}
\Psi_1(r_1,r_2;Y-y)\Psi_2(r_2,r_3;Y-y)r_{13}^4\nabla_1^2\nabla_3^2
\Psi(r_1,r_3;y).
\eeq
As compared to ~\cite{brava} we have a factor $1/4$ since there the
vertex summed all 4 terms in (\ref{vexpr}) which gave identical contribution.
To relate this formula directly to the first term in (\ref{vexpr}) we
introduce coordinates of the 3d gluon and also of the initial pomeron into it.
We also suppress all rapidity dependence together with integration over $y$ which
is of no importance for our purpose. Denoting
\[c\equiv -\frac{1}{4}\frac{g^4N}{4\pi^3}, \ \
F(1,2,3)\equiv\frac{1}{r_{13}^2r_{12}^2r_{23}^2}\]
we then have
\beq
T^{(1)}=c\int\prod_{i=1}^4d^2r_i d^2r'_1d^2r'_2 F(1,2,3)
\delta^2(24)\delta^2(1'1)\delta^2(2'3)\Psi_1(1,2)\Psi_2(4,3)L_{1'2'}
\Psi_3(1'2').
\eeq
In this form it is clear what we get from the rest terms in (\ref{vexpr}).

To find the contribution from the second term in (\ref{vexpr}) we have to
interchange 3$\leftrightarrow$ 4 in the vertex (not touching the wave functions).
We get
\[
T^{(2)}=c\int\prod_{i=1}^4d^2r_i d^2r'_1d^2r'_2 F(1,2,4)
\delta^2(23)\delta^2(1'1)\delta^2(2'4)\Psi_1(1,2)\Psi_2(4,3)L_{1'2'}\Psi_3(1'2')\]\beq=
c\int d^2r_1d^2r_2d^2r_4  F(1,2,4)
\Psi_1(1,2)\Psi_2(4,2)L_{14}\Psi_3(14),
\eeq
or changing integration variable $r_4\to r_3$
\beq
T^{(2)}= c\int d^2r_1d^2r_2d^2r_3  F(1,2,3)
\Psi_1(1,2)\Psi_2(3,2)L_{14}\Psi_3(13)=P_2T^{(1)}.
\eeq
In a similar manner we find the contribution of the third term in (\ref{vexpr}) by
interchanging in $T^{(1)}$ 1$\leftrightarrow $2:
\[
T^{(3)}=c\int\prod_{i=1}^4d^2r_i d^2r'_1d^2r'_2 F(2,4,3)
\delta^2(14)\delta^2(1'2)\delta^2(2'3)\Psi_1(1,2)\Psi_2(4,3)L_{1'2'}\Psi_3(1'2')\]\beq=
c\int d^2r_2d^2r_3d^2r_4F(2,4,3)
\Psi_1(4,2)\Psi_2(4,3)L_{23}\Psi_3(23).
\eeq
Changing integration variables $2\to 1$ and $4\to 2$ we find
\beq
T^{(3)}=
c\int d^2r_1d^2r_2d^2r_3F(1,2,3)
\Psi_1(2,1)\Psi_2(2,3)L_{23}\Psi_3(13)=P_1T^{(1)}.
\eeq
Finally to find the last term coming from (\ref{vexpr}) we have to interchange
both $1\leftrightarrow 2$ and $3\leftrightarrow 4$ in $T{(1)}$:
\[
T^{(4)}=c\int\prod_{i=1}^4d^2r_i d^2r'_1d^2r'_2 F(1,2,4)
\delta^2(13)\delta^2(1'2)\delta^2(2'4)\Psi_1(1,2)\Psi_2(4,3)L_{1'2'}\Psi_3(1'2')\]\beq=
c\int d^2r_1d^2r_2d^2r_4 F(1,2,4)
\Psi_2(4,1)L_{24}\Psi_3(24).
\eeq
Changing $2\to 1, 4\to 3, 1\to 2$ we find
\beq
T^{(4)}=
c\int d^2r_1d^2r_2d^2r_3 F(1,2,3)
\Psi_1(2,1)\Psi_2(3,2)L_{13}\Psi_3(13)=P_1P_2T^{(1)}.
\eeq

Summing all terms we find that the total triple pomeron contribution is
\beq
T=\sum_{i=1}^4T^{(i)}=T^{(1)}(1+P_1+P_2+P_1P_2)=T^{(1)}(1+P_1)(1+P_2).
\eeq
The factor multiplying $T^{(1)}$ is zero if any of outgoing pomerons is
antisymmetric in its gluon coordinates and equal to 4 if they both are
symmetric.
So we find that the pomeron can split only into two symmetric pomerons.
In this derivation it was implicitly assumed that the original pomeron is
symmetric in its gluons (corresponding to its coupling to the $q\bar q$ loop).
It is an open question to see what happens if the original pomeron is
antisymmetric in its gluons.

\section{Appendix 3. Comparing normalization with ~\cite{baryva}(BRV)}
We first compare  normalization of their 3-pomeron vertex $V$.
From Eq. (BRV.74)
we have
\beq
V=\frac{(2N_c)^2g^4}{\sqrt{N_c^2-1}}\frac{\pi^{3/2}}{32}\tilde{V}\simeq
\frac{\pi^{3/2}}{8}g^4N_c\tilde{V},
\eeq
where $\tilde{V}$ is just the Bartels vertex $K_{2\to 3}$ without factors.
On the other hand our definition starts with ~\cite{brava}
\beq
\tilde{Z}D_2=2g^2G=-2g^4N_c K_{2\to3}\otimes D_2,
\eeq
which implies that our vertex
\beq
\Gamma=-g^4N_c\tilde{V}
\eeq
(taking into account that we have to take $1/2$ in view of the fact that $D_2$
is twice the pomeron).
This means that our 3-pomeron vertex is related to BRV as
\beq
V=\frac{\pi^{3/2}}{8}\Gamma.
\eeq

Next we compare normalizations of the Green functions and impact factors.
From (BRV.6) we conclude
\beq
G^{BRV}=2\pi G.
\eeq
The amplitude is in the lowest order
\[A=\frac{is}{2}\int \frac{d^2k}{(2\pi)^3}\Phi_1^{BRV}\Phi^{BRV}_2
\frac{1}{k^2(q-k)^2}=
\frac{is}{2}\int \frac{d^2k}{(2\pi)^2}\Phi_1\Phi_2\frac{1}{k^2(q-k)^2},
\]
wherefrom the relation between the impact factors is
\beq
\Phi^{BRV}=\sqrt{2\pi}\Phi.
\eeq

The BRV impact factor in the conformal representation
\[
\Phi_h^{BRV}=\int \frac{d^3k}{(2\pi)^3}\tilde{E}_h(k,q-k)
\Phi^{BRV}(k,q-k)=\]\[
\int \frac{d^3k}{(2\pi)^3}\Phi^{BRV}(k,q-k)
\int\frac{d^3r_1}{(2\pi)^3}\frac{d^3r_1}{(2\pi)^3}
E_h(r_1,r_1)e^{ikr_1+i(q-k)R-2}\]\[=\frac{\sqrt{2\pi}}{(2\pi)^5}
\int d^2r_1d^2r_2E_h(r_1,r_2)
\int \frac{d^2k_1}{(2\pi)^2}\frac{d^2k_2}{(2\pi)^2}
e^{ik_1r_1+ik_2r_2}(2\pi)^2\delta(k_1+k_1-q)\Phi(k_1,k_2).\]
We conclude from this
\beq
\Phi_h^{BRV}=\frac{\sqrt{2\pi}}{(2\pi)^5}\Phi_h.
\eeq

Now we are in a position to analyze (BRV.(61)). Separating $is/2$ we have
in the $\omega$ representation:
\[
\Delta A=-\frac{(2\pi)^2}{4(2\pi)^4}\Big(\frac{C_{4V}}{16(2\pi)^5}\Big)^2
\]
\[
\int dy_1dy_2\int d\nu d\nu_1 d\nu_2 \nu^2\nu_1^2\nu_2^2
\Phi^{BRV}_h{\Phi^{BRV}_h}^*
g_{Y-y_1,h}g_{y_1-y_2,h_1}g_{y_1-y_2,h_2}g_{y_2,h}=
\]
\[
-\frac{2\pi}{(2\pi)^{10}}\frac{(2\pi)^2}{4(2\pi)^4}
\Big(\frac{C_{4V}}{16(2\pi)^5}\Big)^2
\]\[
\int dy_1dy_2\int d\nu d\nu_1 d\nu_2 \nu^2\nu_1^2\nu_2^2
\Phi_h{\Phi_h}^*
g_{Y-y_1,h}g_{y_1-y_2,h_1}g_{y_1-y_2,h_2}g_{y_2,h}
\]
where $h=1/2+i\nu$, $h_{1(2)}=1/2+i\nu_{1(2)}$ and
\[ C_{4V}=2^{12}\pi^{7/2}\alpha_s^2N
\Omega\left(\frac{1}{2},\frac{1}{2},\frac{1}{2}\right).
\]
Since
$\nu^2=\pi^4/(2a_h)$,
we find
\[
\Delta A=
-\frac{2\pi}{(2\pi)^{10}}\frac{(2\pi)^2}{4(2\pi)^4}
\Big(\frac{C_{4V}}{16(2\pi)^5}\Big)^2
\pi^{12}
\]\[
\int dy_1dy_2\int \frac{d\nu}{2a_h}\frac{d\nu_1}{2a_{h_1}}
\frac{d\nu_2}{2a_{h_2}} \Phi_h{\Phi_h}^*
g_{Y-y_1,h}g_{y_1-y_2,h_1}g_{y_1-y_2,h_2}g_{y_2,h},
\]
which implies
\beq
\Sigma_{h,y_1-y_2}^{BRV}
=C\int \frac{d\nu_1}{2a_{h_1}}
\frac{d\nu_2}{2a_{h_2}}
g_{y_1-y_2,h_10}g_{y_1-y_2,h_2}
\eeq
where
\beq
C=\frac{2\pi}{(2\pi)^{10}}\frac{2\pi)^2}{4(2\pi)^4}
\Big(\frac{C_{4V}}{16(2\pi)^5}\Big)^2
\pi^{12}
=\frac{\alpha_s^4N^2}{\pi^2}
\Omega^2\left(\frac{1}{2},\frac{1}{2},\frac{1}{2}\right).
\eeq
Comparing with our expression we find
\beq
\Sigma_{0\nu}^{BRV}=\frac{1}{8}\Sigma_{0\nu}.
\eeq

\end{document}